\documentclass[a4paper,fleqn,usenatbib]{mnras}

\usepackage[T1]{fontenc}
\usepackage{ae,aecompl}

\usepackage{graphicx}	
\usepackage{amsmath}	
\usepackage{amssymb}	

\newcommand{\ser}{S\'ersic}

\title[PRGs candidates]{New candidates to polar-ring galaxies from the Sloan
Digital Sky Survey}

\author[V. P. Reshetnikov, A. V. Mosenkov]{
Vladimir P. Reshetnikov,$^{1}$
Aleksandr V. Mosenkov,$^{2}$\thanks{E-mail: v.reshetnikov@spbu.ru, mosenkovav@gmail.com}
\\
$^{1}$St.Petersburg State University, 7/9 Universitetskaya nab., St.Petersburg, 199034 Russia
\\
$^{2}$Central Astronomical Observatory, Russian Academy of Sciences, 65/1 Pulkovskoye chaussee, St. Petersburg, 
196140 Russia 
}

\date{Accepted 2018. Received 2018; in original form 2018}

\pubyear{2018}

\begin{document}
\label{firstpage}
\pagerange{\pageref{firstpage}--\pageref{lastpage}}
\maketitle

\begin{abstract}
Polar-ring galaxies (PRGs) are fascinating systems in which the
central object (typically, an early-type galaxy) is encircled by a large-scale
ring of stars, gas, and dust with almost perpendicular spin. PRGs
are rare objects and their formation mechanism is not entirely clear.
In this paper, we present a new sample of 31 candidates to PRGs 
identified in the Sloan Digital Sky Survey (SDSS). Using their stacked $gri$
images, we determined geometrical parameters of these galaxies
(the position angle and the size of the host galaxy and the ring).
We compare our sample objects to the previously known PRGs and discuss
their general characteristics (the frequency of faint outer structures, 
the luminosity--size relation, and the distribution by the apparent angle between 
the ring and the host galaxy). Our main results are:
(i) Central galaxies of PRGs follow the luminosity--size relation for ordinary
galaxies. The ring components are located along the similar relation but with
a larger scatter. This increasing scatter may be the result of a secondary origin
of polar structures. (ii) At least half of PRGs show a ring component within
20$^\circ$ from the perpendicular orientation.

\end{abstract}




\section{Introduction}

Polar-ring galaxies (PRGs) provide remarkable examples of the non-monotonicity 
of the galaxy formation process. Typically, they consist of two large-scale 
luminous subsystems: a central galaxy and a ring or a disc which are oriented 
nearly orthogonally to the major axis of the central object (see examples in the
\citealt{whit1990} (= PRC) and \citealt{moiseev2011} (= SPRC) catalogues).
In this article, we further use the term ``PRG'' to designate a 
class of objects
with near-polar optical structures -- without dividing into polar-disc or polar-ring
galaxies (see, for instance, \citealt{iodice2014}).

In most cases, two subsystems in PRGs differ noticeably by their characteristics.
The central (host) galaxies are typically early-type (E/S0) galaxies, poor in gas
(\citealt{wms1987}; \citealt{ffb2012}; \citealt{rc2015}).
In contrast, the polar structures are generally younger, blue 
(e.g., \citealt{rhy1994}; \citealt{iod2002}), 
with large amount (several times 10$^9$\,M$_{\odot}$) of HI 
(e.g., \citealt{vg1987}; \citealt{vdr2000}). 

Various scenarios have been proposed to explain the formation of the two 
kinematically and morphologically decoupled structures in PRGs:
the capture of matter from an approached galaxy (\citealt{reshsotn1997,bourcomb2003}), 
the merging of galaxies (\citealt{bekki1997,bekki1998,bourcomb2003}), 
the misaligned accretion of matter from cosmic filaments 
(e.g. \citealt{maccio2006,brook2008}), and some others. Observations show that 
several formation mechanisms can apparently be realised.

Galaxies with polar structures are an important class of objects that allows us
to study a wide range of problems, linked with the formation and evolution 
of galaxies under external accretion, and to study the properties of their
dark haloes. Unfortunately, PRGs are very rare objects, with a fraction 
of $\sim10^{-3}$  of all galaxies in the local Universe 
(\citealt{whit1990, resh2011}). Consequently, most recent studies
of PRGs are concentrated mainly on a few relatively nearby and bright galaxies:
NGC\,4650A (e.g., \citealt{cia2014}, \citealt{icc2015}), A0136-0801
(\citealt{sia2015}), NGC\,660 (\citealt{swm2017}), and some others.

There were several attempts to collect large samples of such galaxies.
\citet{whit1990} have presented a photographic
atlas of PRGs and related objects containing 157 galaxies
but, according to further studies, only a small part of the atlas galaxies 
appeared to be true PRGs. 

The next important step was the catalogue SPRC by \citet{moiseev2011} 
based on the results from the Galaxy Zoo project\footnote{http://www.galaxyzoo.org/} \citep{2008MNRAS.389.1179L}. The SPRC
catalogue contains 275 objects split into four groups. The first
two groups comprise ``best candidates'' (70 objects) and
``good candidates'' (115 objects) for PRGs. The remaining objects are
related galaxies (mergers, strong warps) and possible
face-on polar rings. A further study showed that an appreciable fraction 
of the SPRC objects are indeed PRGs (fig.\,1 in \citealt{moiseev2015} gives 
several examples).

In this work we present a new sample of ``best'' and ``good'' (in the sense of
SPRC) candidates to PRGs. Our sample includes 31 galaxies compiled 
from Galaxy Zoo discussion boards. 

Throughout this article, we adopt a standard flat $\Lambda$CDM
cosmology with $\Omega_m$=0.3, $\Omega_{\Lambda}$=0.7, 
$H_0$=70 km\,s$^{-1}$\,Mpc$^{-1}$. All magnitudes in the paper are given
in the AB-system.

\section{THE SAMPLE AND DATA REDUCTION}

\subsection{The sample}
\label{sample}

To select new candidates to PRGs, we analysed discussion boards\footnote{https://talk.galaxyzoo.org/\#/boards}  
of the Galaxy Zoo project. By eye looking
at the morphology of these galaxies (using coloured SDSS snapshots), we finally 
selected a sample of 31 objects with a morphology resembling 
``classical'' PRGs, like A0136-0801, NGC\,4650A, NGC\,2685 (see \citealt{whit1990} 
catalogue). The selected objects show two inclined large-scale structures, 
the centres of which nearly coincide.
The full list of galaxies, along with some of their general properties, which are derived 
in Sect.~\ref{analysis}, is presented in Table~\ref{tab1}. The reduced 
images of all candidates to PRGs are shown in the Appendix~\ref{red_images}, 
Fig.~\ref{images} (see the description of these images below). 

We should notice that our selection is strongly biased 
due to the fact that only PRGs with highly-inclined and extended polar 
structures (PS) can be
easily found. As mentioned by \citet{whit1990}, PS that has a smaller
diameter than the host galaxy (HG), i.e. a narrow polar-ring, and that has a small
inclination angle with respect to the HG are difficult to detect.
This selected sample of galaxies cannot be complete by any means, and can 
only be considered as an extension of the already existing sample of candidates 
to this rare class of PRGs \citep[see e.g.][]{moiseev2011}.

Most objects in our sample look typical for PRGs -- they possess an extended
edge-on component which is almost orthogonal to the central galaxy. In two cases,
the galaxies \#\,15 (SDSS\,112850.42-061614.1) and \#\,16 
(SDSS\,124513.41-154102.0), the suspected 
PSs are moderately inclined to the line-of-sight. The galaxy \#\,15
(SDSS\,112850.42-061614.1) looks very similar to the kinematically-confirmed 
polar-ring galaxy SPRC-7 (\citealt{brosch2010}). 

\begin{table*}
 \centering
 \begin{minipage}{180mm}
  \centering
  \parbox[t]{150mm} {\caption{General characteristics of the sample galaxies. 
  The columns 4--11 are described in Sect.~\ref{reduction}--\ref{analysis}.}
  \label{tab1}}
  \begin{tabular}{ccccccccccc}
  \hline 
  \hline
  \# &  SDSS  & $z$ & $r_\mathrm{as}$           & $M^0_r$ & $(g - r)_0$ & $R_\mathrm{2\sigma}$ & $R_\mathrm{eff}$ & $\Delta$P.A. & $D_\mathrm{r}$ & $D_\mathrm{r}/D_\mathrm{h}$ \\ 
    &  name  &      & (mag)         &         &  & (kpc) & (kpc)             &  (grad)      & (kpc) & \\
  (1) & (2) & (3) & (4) & (5) & (6) & (7) & (8) & (9) & (10) & (11) \\
\hline 
1 & 001524.31+184624.8 & 0.0186 &  $14.35\pm0.02$ & -20.25 & 0.72 & 9.2 & 3.5 & 86 & 18.4 & 1.20\\
2 & 004914.26+120858.1 & 0.0383 &  $14.57\pm0.05$ & -21.71 & 0.77 & 12.8 & 2.6 & 81 & 24.8 & 1.00\\
3 & 010816.84-082317.6 & 0.0467 &  $15.08\pm0.03$ & -21.60 & 0.82 & 14.8 & 9.2 & 78 & 29.6 & 1.15\\
4 & 013905.96+192559.9 & 0.0880 &  $16.38\pm0.05$ & -21.78 & 0.72 & 13.2 & 4.8 & 87 & 26.4 & 1.16\\
5 & 015828.87+013549.9 & 0.0600 &  $16.60\pm0.06$ & -20.59 & 0.69 & 8.4 & 2.7 & 68 & 14.4 & 0.86\\
6 & 020316.58+293830.4 & (0.038) & $14.83\pm0.04$ & -- & 0.72 & -- & -- & 79 & -- & 0.45\\
7 & 041557.96-051655.8 & 0.0317 &  $15.08\pm0.03$ & -20.83 & 0.86 & 9.8 & 4.6 & 86 & 19.5 & 1.43\\
8 & 074308.88+240206.5 & 0.0880 &  $16.88\pm0.03$ & -21.27 & 0.66 & 26.7 & 8.1 & 88 & 53.4 & 4.06\\
9 & 075131.01+500216.7 & 0.0600 &  $16.37\pm0.08$ & -20.92 & 0.77 & 11.5 & 4.0 & 80 & 22.2 & 1.05\\
10 & 081740.08+042952.3 & 0.1100 & $17.01\pm0.03$ & -21.66 & 0.82 & 14.3 & 6.4 & 81 & 28.5 & 1.26\\
11 & 091522.74+271711.9 & 0.0460 & $14.28\pm0.02$ & -22.28 & 0.63 & 26.7 & 11.1 & 69 & 53.4 & 1.11\\
12 & 091558.60+093455.3 & 0.0450 & $15.92\pm0.05$ & -20.67 & 0.79 & 12.8 & 2.1 & 78 & 25.6 & 1.82\\
13 & 105157.88+085122.0 & 0.0520 & $14.67\pm0.02$ & -22.20 & 0.71 & 19.6 & 11.6 & 84 & 27.9 & 0.71\\
14 & 112301.31+470308.7 & 0.0250 & $14.32\pm0.03$ & -20.86 & 0.75 & 9.0 & 3.5 & 85 & 18.0 & 1.07\\
15 & 112850.42-061614.1 & 0.0202 & $15.85\pm0.03$ & -18.89 & 0.43 & 10.4 & 2.7 & 24 & 20.7 & 2.88\\
16 & 124513.41-154102.0 & 0.0440 & $14.76\pm0.03$ & -21.76 & 0.63 & 18.8 & 6.4 & 70 & 37.6 & 1.52\\
17 & 133040.07+113541.7 & 0.0240 & $15.05\pm0.02$ & -20.07 & 0.67 & 10.1 & 5.6 & 83 & 15.7 & 0.77\\
18 & 135711.99-070430.3 & 0.0275 & $14.17\pm0.01$ & -21.28 & 1.03 & 8.4 & 11.4 & 52 & 7.6 & 0.45\\
19 & 140435.21+555536.8 & (0.128) &$18.22\pm0.09$ & -- & 0.81 & -- & -- & 78 & -- & 1.91\\
20 & 161103.94+140043.6 & 0.0310 & $15.86\pm0.05$ & -19.86 & 0.62 & 6.5 & 2.5 & 64 & 12.9 & 1.28\\
21 & 162434.00+364015.5 & 0.0500 & $16.17\pm0.08$ & -20.59 & 0.80 & 7.4 & 1.8 & 73 & 11.1 & 0.75\\
22 & 165057.26+044931.4 & (0.071) &$16.54\pm0.01$ & -- & 0.52 & -- & -- & 61 & -- & 0.90\\
23 & 175659.49+040143.3 & (0.117) &$17.47\pm0.06$ & -- & 0.60 & -- & -- & 67 & -- & 1.02\\
24 & 204425.23+022322.2 & 0.0595 & $16.28\pm0.04$ & -21.06 & 0.56 & 13.1 & 5.2 & 63 & 17.6 & 0.67\\
25 & 205213.48+034924.3 & (0.034) &$16.22\pm0.06$ & -- & 0.38 & -- & -- & 76 & -- & 0.97\\
26 & 211245.01+092938.0 & 0.0960 & $17.05\pm0.06$ & -21.46 & 0.67 & 13.5 & 3.7 & 72 & 27.2 & 1.30\\
27 & 220419.41+125806.2 & 0.0270 & $14.25\pm0.03$ & -21.20 & 0.79 & 11.6 & 3.6 & 74 & 12.9 & 0.56\\
28 & 221126.92+175348.4 & 0.0200 & $16.60\pm0.04$ & -18.16 & 0.77 & 4.3 & 1.5 & 53 & 8.6 & 1.78\\
29 & 224835.75+072424.8 & (0.082) &$16.74\pm0.06$ & -- & 0.72 & -- & -- & 90 & -- & 0.99\\
30 & 234407.32+253657.7 & 0.0960 & $16.00\pm0.03$ & -22.37 & 0.80 & 18.6 & 8.8 & 88 & 37.2 & 1.50\\
31 & 234756.34+080151.4 & (0.042) &$15.00\pm0.02$ & -- & 0.37 & -- & -- & 73 & -- & 0.62\\

  \hline\\
  \end{tabular}

\end{minipage}
   
   \parbox[t]{160mm}{ Columns: \\
   (1) designation number in the sample, \\
   (2) SDSS name,  \\
   (3) spectroscopic redshift (SDSS or NED), brackets mark photometric 
           redshifts (SDSS),  \\
   (4) asymptotic magnitude in the $r$ band, \\
   (5) absolute magnitude in the $r$ band corrected for the Milky Way 
           extinction \citep{schfin2011} and $k$-correction \citep{chil2010},  \\ 
   (6) $g - r$ color calculated from the corresponding asymptotic magnitudes and corrected for the Milky Way extinction \citep{schfin2011} 
           and $k$-correction \citep{chil2010},  \\ 
   (7) radius of the $2\sigma$-isophote estimated for the stacked $gri$-image, \\
   (8) effective radius in the $r$ band, \\
   (9) apparent angle between the ring and the central galaxy,  \\
   (10) diameter of the suspected ring,  \\
   (11) the ring diameter normalised by the diameter of the host galaxy.   }
   \end{table*} 

\subsection{The data reduction}    
\label{reduction}
    
To highlight faint polar structures, which may be present in galaxies, 
we used the following technique adopted from \citet{2011A&A...536A..66M}.
The system response curves of the SDSS imaging camera \citep{gunncam1998} 
show that the highest values are reached in the filters $g$, $r$ and $i$, 
therefore we analysed the galaxy images in these three filters only. The corrected 
frames were directly downloaded from the SDSS archive, 
Data Release 12 \citep{2015ApJS..219...12A}. Most of the galaxies lie within 
one frame, but for those galaxies, which lie at the border of the frame, 
additional overlapping frames were downloaded and combined together using 
the {\sc swarp} routine \citep{2002ASPC..281..228B}. 

Then we determined the radius $R_{2\sigma}$ of the circle which encompasses 
the $2\sigma$-level isophote of the stacked $gri$-image (see below) of the galaxy, 
which was initially created from the directly downloaded images in the $g$, $r$ and $i$ bands (no sky-background correction). 
After that, we created an initial mask using \textsc{sextractor} 
\citep{1996A&AS..117..393B} and applied  sigma-clipping to estimate the median 
and standard deviation of the background in each band. Although SDSS frames have 
been already sky-subtracted, we decided to re-estimate the background near each 
galaxy. For this purpose, to each galaxy frame we applied the \textsc{iraf} 
\textit{ellipse} routine to create the growth curve. Then we defined an annulus 
in which the background was carefully estimated (again, using sigma-clipping) 
as having the inner radius $1.50\,R_{2\sigma}$ and the outer radius 
$1.80\,R_{2\sigma}$. The area of the annulus in this case is equal to the area 
inside the circle which is circumscribed about the galaxy. We double checked 
that within this annulus the growth curve is flat. We did not apply the polynomial 
fitting of the sky background because the SDSS frames have been already 
sky-subtracted and we found no gradient of the sky background. The calculated 
background (which appeared to be within 1\% of the background level which was 
determined in the SDSS) was then subtracted from each frame.

Asymptotic magnitudes in each band were estimated by extrapolating the 
dependence of the gradient $\mathrm{d}m/\mathrm{d}r$ on magnitude to 
$\mathrm{d}m/\mathrm{d}r=0$ (by definition, this gives the asymptotic magnitude 
of the galaxy, see e.g. \citealt{2015ApJS..219....3M}). The estimated 
uncertainties in the magnitudes are a combination of the associated sky error 
value, the Poisson error on the incident flux, and the the SDSS calibration 
error of 0.8\% \citep{2015ApJS..219...12A}. In addition, we also found the 
effective radius of each galaxy, using the created growth curves.

The obtained images in each filter were stacked together using the \textsc{iraf} 
task \textit{imcombine}. 
This was done in order to improve the signal-to-noise ratio of each individual 
image. After that, for all three filters we created individual mask images
which contain contaminating stars and galaxies. This step was done using 
the segmentation maps, produced by {\sc sextractor}, which 
were then carefully revisited by eye to ensure that all foreground stars 
and galaxies had been removed from the galaxy images. However, we decided 
not to mask out close, overlapping galaxies (as in the case of 
SDSS\,J135711.99-070430.3) since the light from them non-negligibly affects 
the surface brightness distribution of the polar-ring galaxy. 
After that, to mask all contaminating pixels in the stacked galaxy image, 
the masks in all three filters were combined. 

In the next step, to make faint features visible, we used a circularly 
symmetric Gaussian filter with $\sigma = 2-4$~pix; these values were found 
optimal for each galaxy to achieve a good enhancement of a faint ring 
structure. Since in the vicinity of the sample galaxies no bright stars 
are found on the final images (except for SDSS\,J105157.88+085122.0) and the 
galaxies are rather small, we did 
not found large gradients of the background. The radius $R_{2\sigma}$ was 
re-determined for the final $gri$-images. 

\begin{figure}
\includegraphics[width=8.5cm, angle=0, clip=]{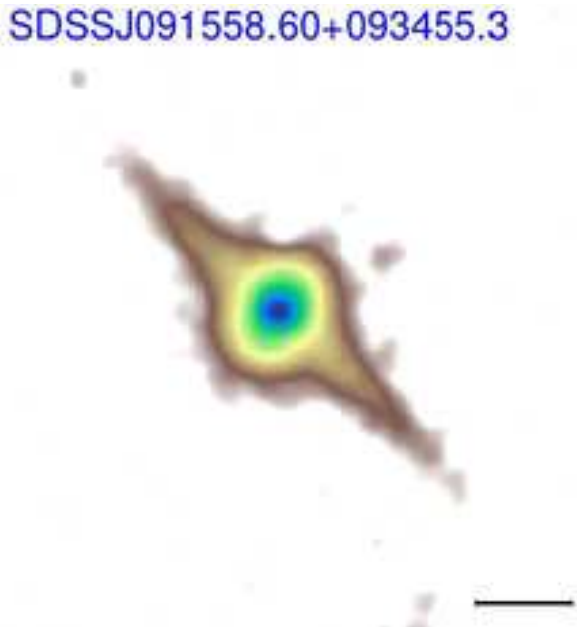}
\caption{Reduced (stacked) $gri$-image of SDSS\,J091558.60+093455.3. The 
contaminating sources have been masked. The scale bar corresponds to 10\arcsec.}
\label{ex_image}
\end{figure}

In Fig.~\ref{ex_image} we show a typical PRG in our sample, SDSS\,J091558.60+093455.3, 
whereas $gri$-images for all sample galaxies are given in the Appendix in Fig.~\ref{images}.  
As can be seen, in some galaxies a PS is well prominent, whereas in others it is 
quite faint and barely visible.

\subsection{The analysis of the images}    
\label{analysis}

To estimate the apparent position angle of the HG and PS, 
we applied the following approach. Since in most galaxies in our sample the 
PS looks as a rather faint almost orthogonally oriented structure, 
the proper way to determine the orientation of the host galaxy and the 
polar structure is by fitting the whole galaxy with a two-component model `host + polar structure'.
For this purpose, we used a new version of the \textsc{deca} 
package \citep{2014AstBu..69...99M} which was specially designed to perform 
automatic and semi-automatic decomposition of galaxy images onto several 
structural components. Most galaxies in our sample consist of a HG and a PS which can be both generally described 
by a \ser\ function \citep{ser1968}. This approximation is justified 
by the following reasons. First, the sample galaxies are rather small, 
and, thus, are poorly resolved to be properly fitted with a more complex 
model. Second, the polar structure in most galaxies from our 
sample is oriented in a way that no contrast ring is seen. But even in the case of SDSS\,J112850.42-061614.1 (where the PS is moderately inclined towards the observer and seen well), 
a model with a very small \ser\ index ($n<0.1$) strives to fit the ring 
structure by making the central part of the model flat. This method has been 
tested successfully on a sample of galaxies with apparent ring structures in 
the far-infrared (Mosenkov et al., submitted). Also, since 
our aim is to merely estimate the position angles of almost orthogonal 
components, this rough fitting scheme allows us to do this very easily. 
Another way to find the position angles of the HG and PS would be using 
the distribution of the position angle by radius, which has been derived 
using the \textsc{iraf} task \textit{ellipse}. However, we found that since 
most PS in our sample are barely visible, this approach does not allow us 
to robustly determine the difference in the position angles at different radii, 
and, thus, clearly discern the HG and PS in the distributions of the position 
angle by radius.

In some galaxies additional components (apart from the HG and PS) were fitted with 
another \ser\ function to create a realistic model of the HG and PS. 
For SDSS\,J135711.99-070430.3, 
an additional \ser\ function was used to describe an interacting galaxy. 

\begin{figure*}
\includegraphics[width=18cm, angle=0, clip=]{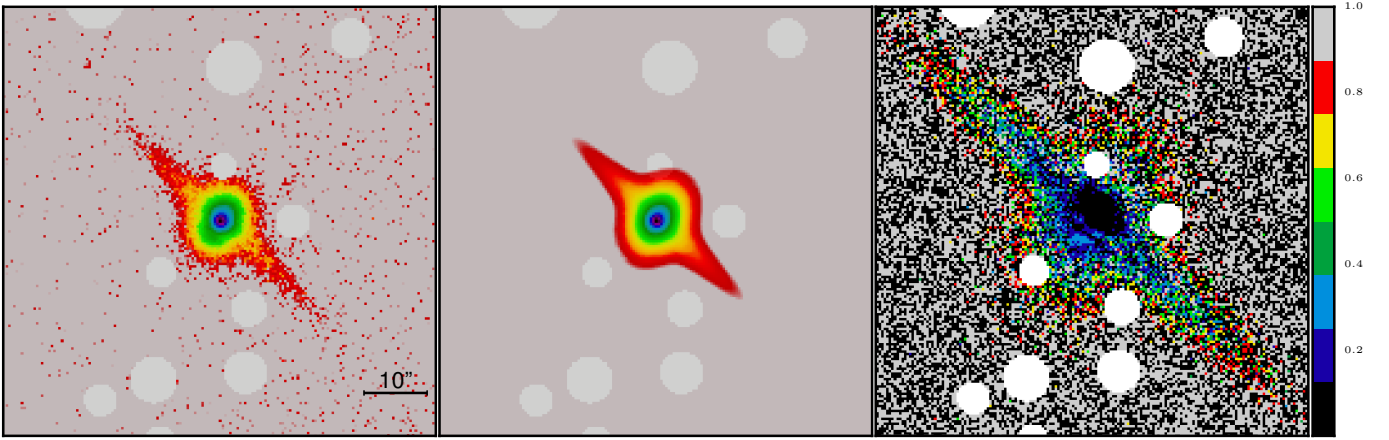}
\caption{
The results of the fitting for SDSS\,J091558.60+093455.3 for a stacked $gri$-image: the stacked $gri$-image (left panel), the best fitting image (middle panel),
and the residual image (right panel) which indicates the relative deviation between
the fit and the image in absolute values, i.e. $| data-model |/model$. The right color bar shows the scaling of the residual image. The masked objects are highlighted by the white colour. The galaxy and model images are given in a logarithmic scale. All images cover a field-of-view of $1\arcmin\times1\arcmin$.}
\label{ex_decomp}
\end{figure*}

Finally, we successfully applied \textsc{deca} to the sample galaxies, 
using the created stacked $gri$-images and their masks. In Fig.~\ref{ex_decomp} 
we show the results of the fitting for SDSS\,J091558.60+093455.3. The difference of 
the apparent position angle of the polar structure and the host galaxy ($\Delta$P.A.), 
which were retrieved from the \textsc{deca} fitting, are presented in Table\,1. 

\begin{figure}
\includegraphics[width=8.5cm, angle=0, clip=]{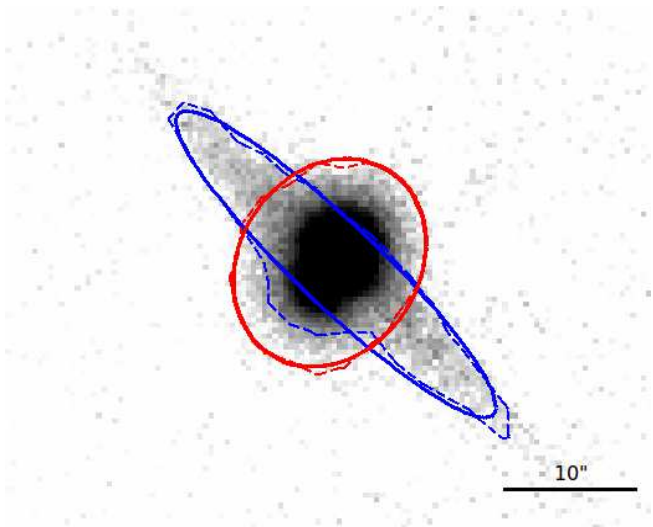}
\caption{The results of the ellipse fitting for a stacked $gri$-image of 
the galaxy SDSS\,J091558.60+093455.3. The dashed contours are the outermost 
isophotes (at $2\sigma$), which were created for the residual images 
`galaxy - host' and `galaxy - ring'. The red solid contour shows the fit ellipse
for the HG contour and the blue solid contour represents the fit ellipse
for the PS contour.}
\label{example_iso}
\end{figure}

In addition, we provide the diameters of the host galaxy $D_\mathrm{HG}$ and 
the polar structure $D_\mathrm{PS}$ which were derived from the residual image 
`galaxy - ring structure' and `galaxy - host', respectively. The outermost isophotes 
of each residual image were plotted at the $2\sigma$ background level 
(which corresponds to a surface brightness of $\mu_r\approx25.5-26$~mag\,arcsec$^{-2}$) 
and then fitted with ellipses with the major axes $D_\mathrm{PS}$ and $D_\mathrm{HG}$ 
(see Fig.\,\ref{example_iso}). 

\begin{figure*}
\centering
\includegraphics[width=5cm, angle=0, clip=]{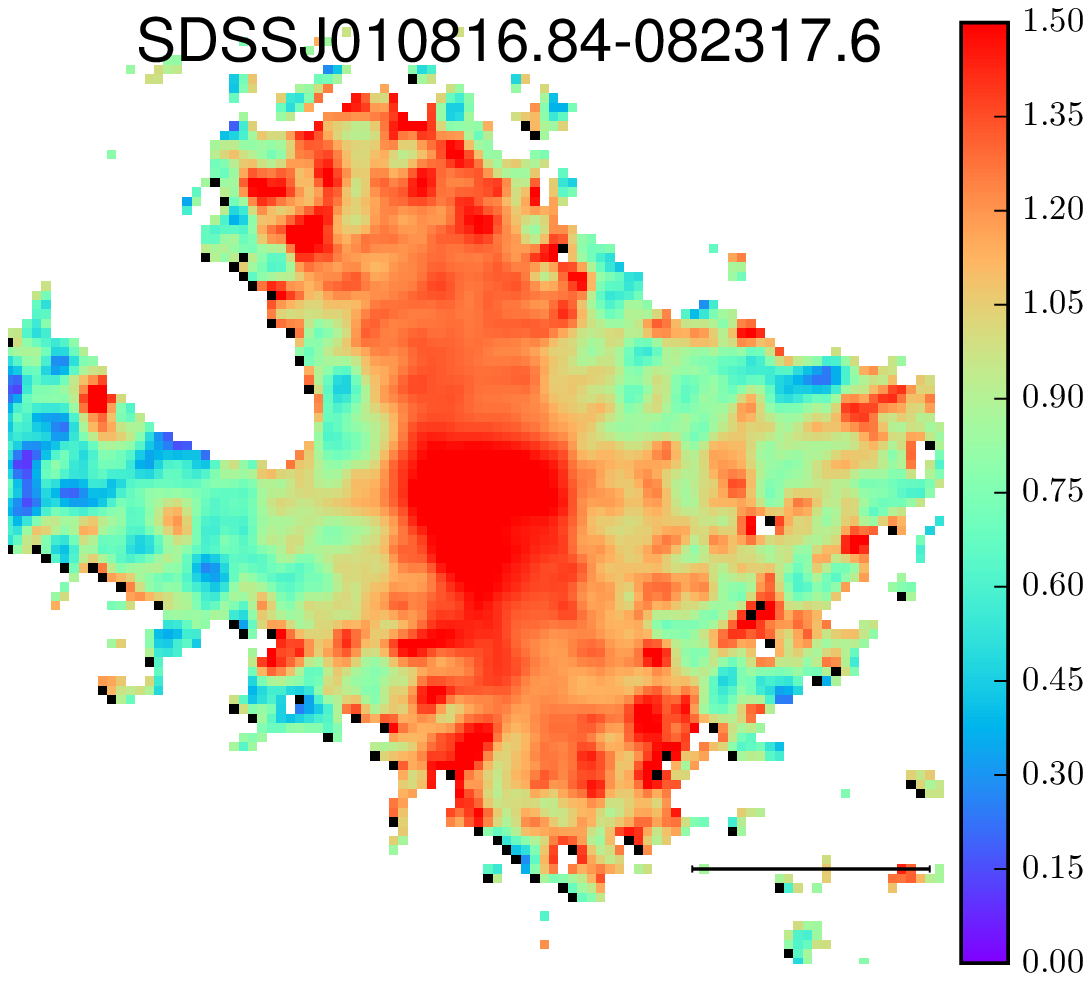}
\includegraphics[width=5cm, angle=0, clip=]{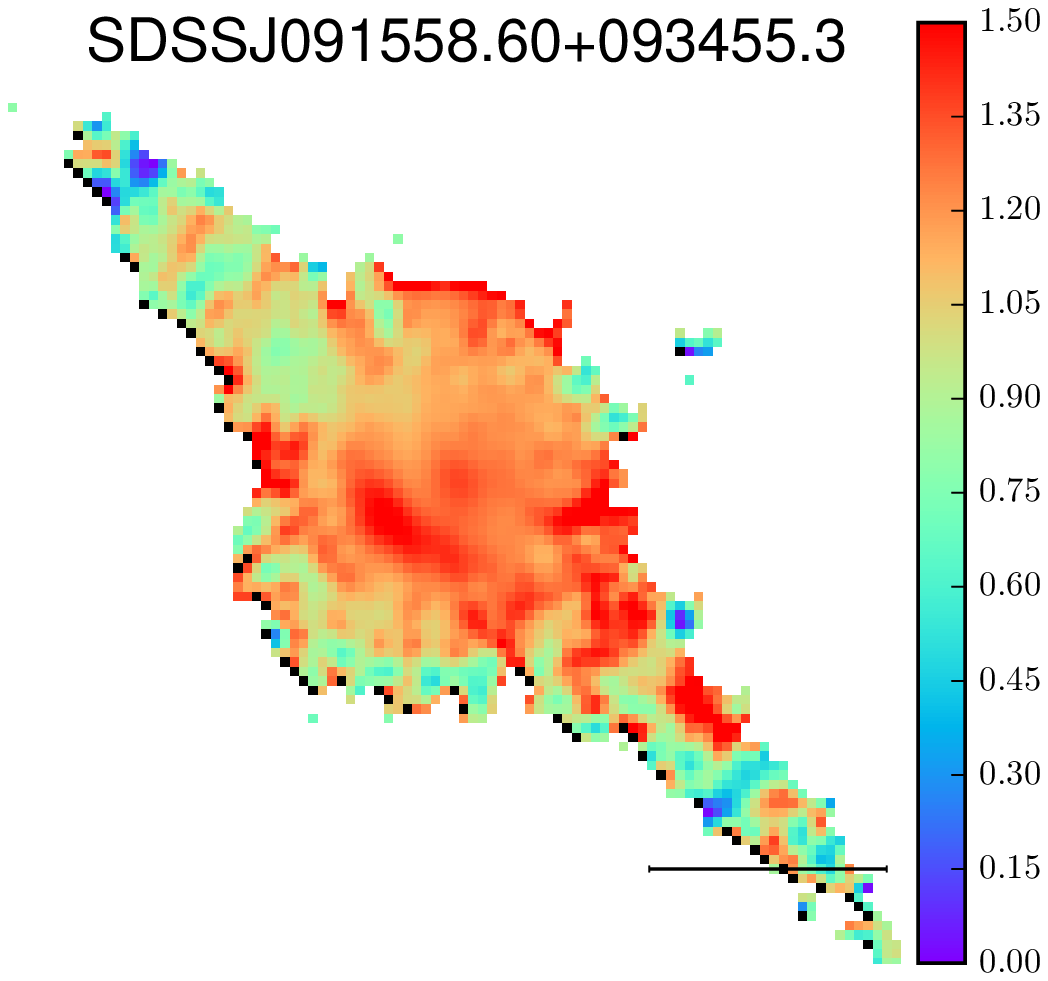}
\includegraphics[width=5cm, angle=0, clip=]{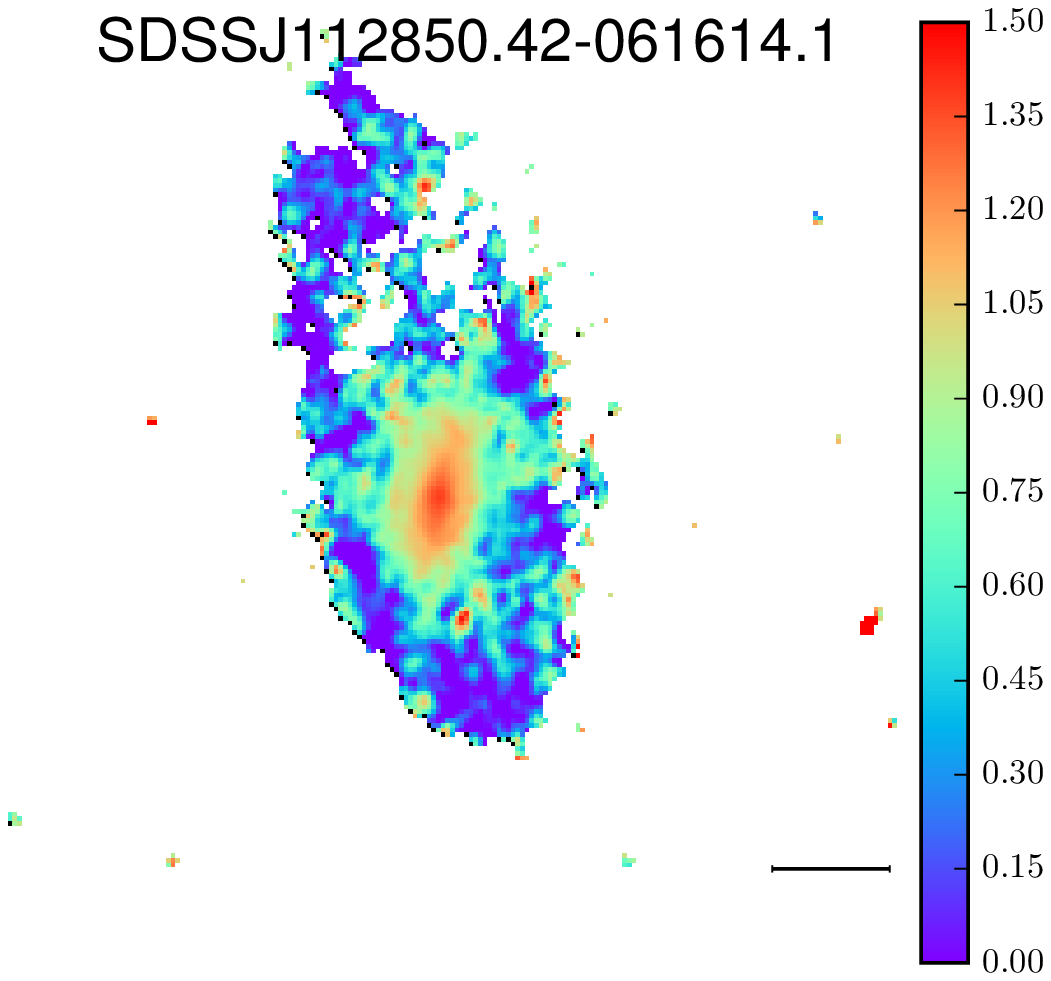}
\includegraphics[width=5cm, angle=0, clip=]{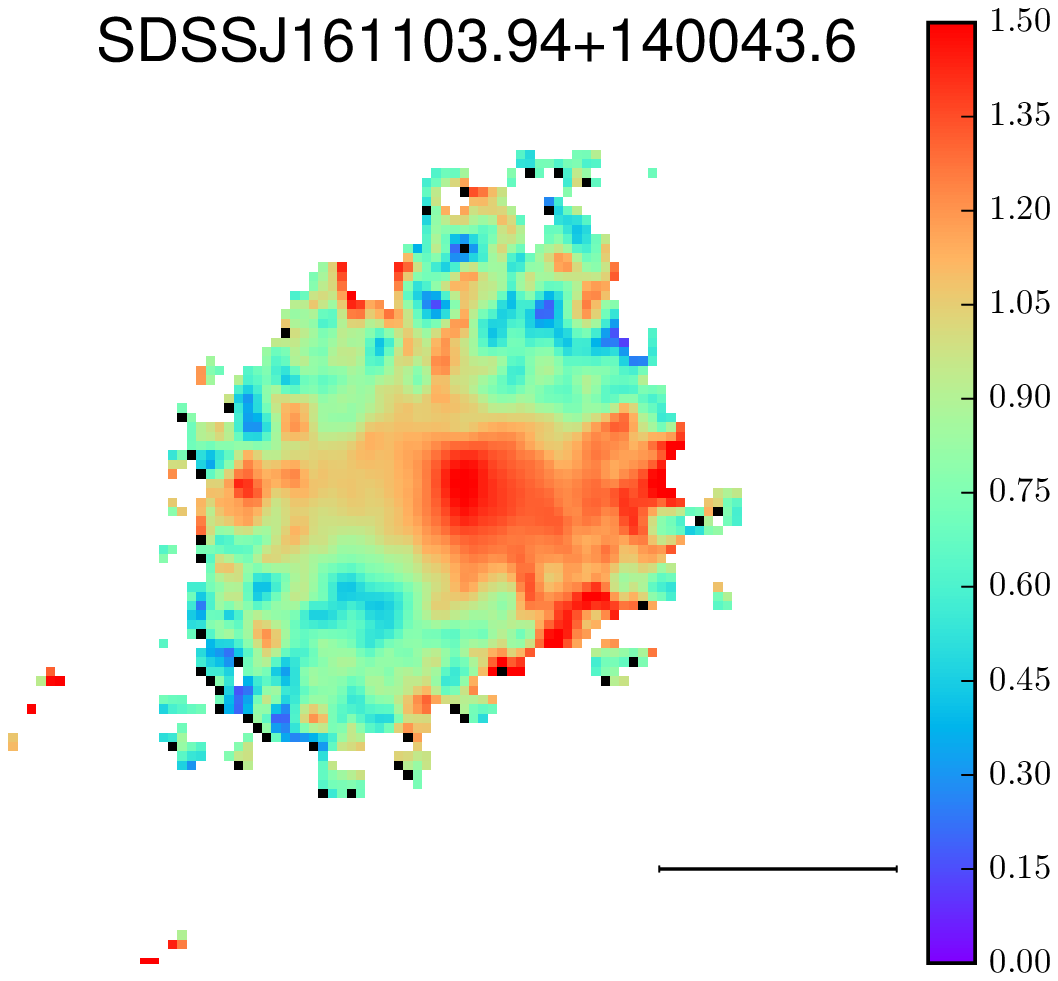}
\includegraphics[width=5cm, angle=0, clip=]{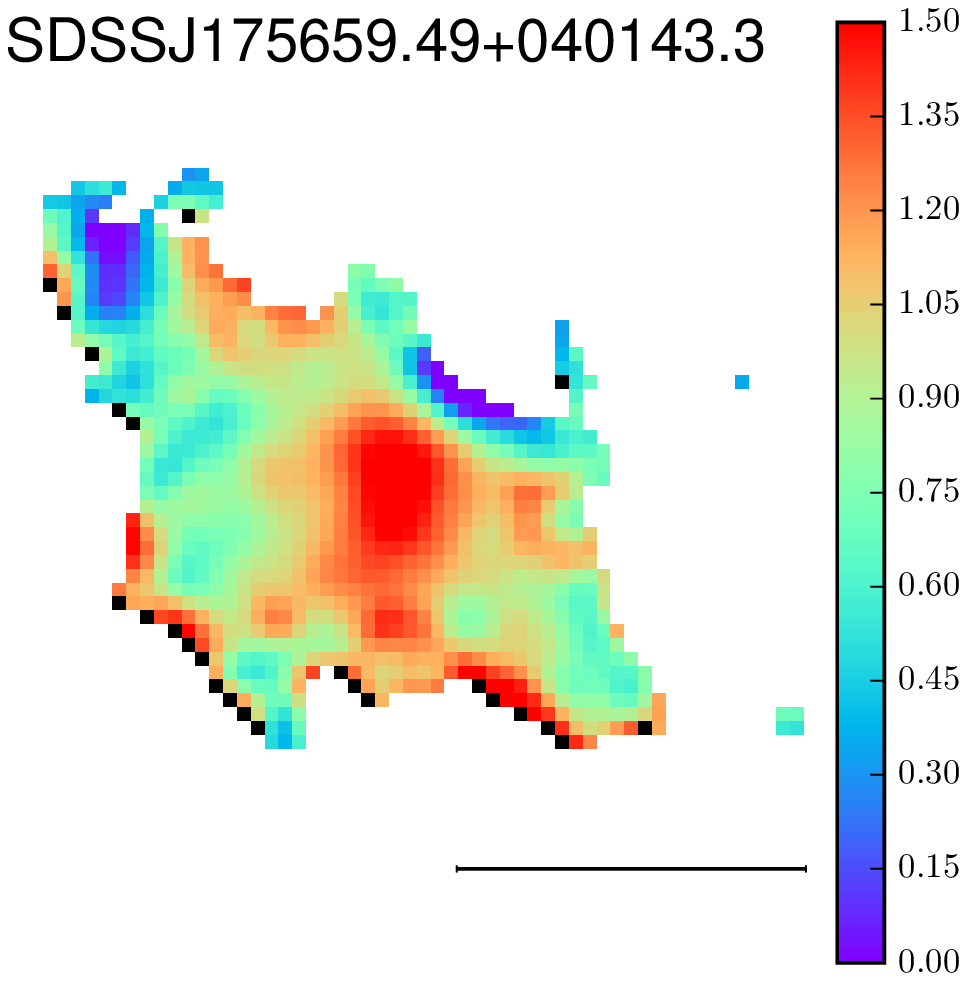}
\includegraphics[width=5cm, angle=0, clip=]{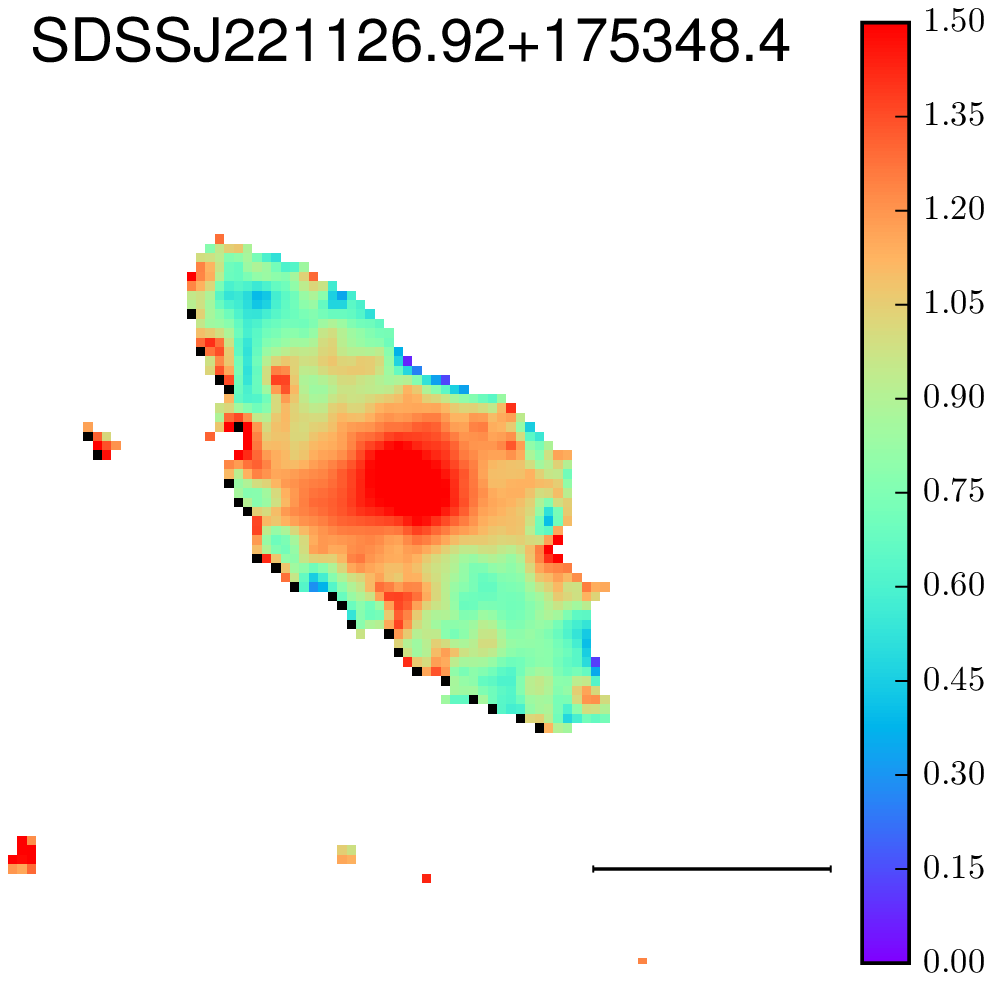}
\caption{The $g - i$ color maps for 6 candidates to PRGs. The images are limited to
a $2\sigma$-level isophote created for their corresponding $gri$-images. The scale 
bar in each plot corresponds to 10\arcsec.}
\label{cols_maps}
\end{figure*}

Fig.\,\ref{cols_maps} provides the $g - i$ color maps for several PRG candidates.
Generally, the sample galaxies are faint and small. Thus, in most cases their 
color maps are irregular and noisy. We show in Fig.\,\ref{cols_maps} some
best examples. As one can see in the figure, the HGs look rather red while the
PSs are generally bluer.

\section{DISCUSSION}

\subsection{General characteristics of PRGs}
\label{char}

In this section we compare characteristics of our sample galaxies with
that of \citet{smirmois2013}. \citet{smirmois2013} presented the
combined sample of ``best candidates'' from the SPRC (70 galaxies) and of
8 kinematically-confirmed PRGs from the SPRC and PRC catalogues with
available SDSS images. The authors carried out measurements of the shape,
sizes, and position angles of the host galaxy and the polar structure.
At present, this is the largest published sample with 
the homogeneous estimates of the PRGs geometrical characteristics.
Our method for measuring the PRGs sizes and position angles 
(see previous section) provides close results compared to those from 
\citet{smirmois2013}.

\begin{figure*}
\includegraphics[width=11.0cm, angle=-90, clip=]{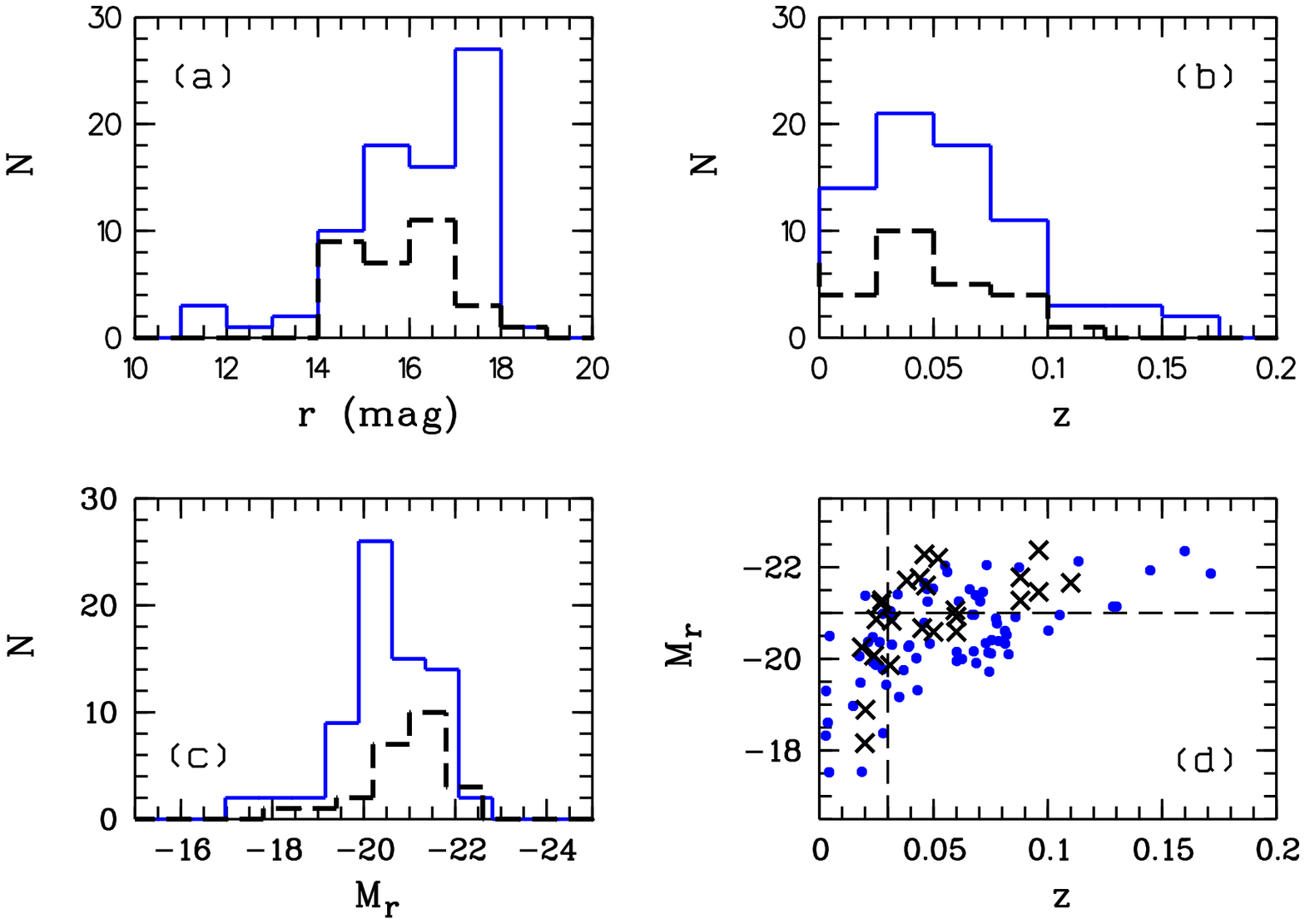}
\caption{Histograms of the distributions of PRG candidates by
(a) apparent $r$-band magnitude, (b) redshift, (c) absolute magnitude 
for the \citet{smirmois2013} (the blue solid line) and our (the dashed line) sample.
Panel (d) shows the dependence of absolute magnitude of PRGs on their
redshift (the blue points -- data from \citet{smirmois2013}, the crosses -- data
from Table~\ref{tab1}). The dashed lines separate the two subsamples (see the text).}
\label{char1}
\end{figure*}

In Fig.~\ref{char1} (panels \textit{a--c}), we show the comparison of the distributions 
by the apparent magnitude, the redshift, and the absolute magnitude
of galaxies from \citet{smirmois2013} and from our Table~\ref{tab1}.
As one can see, both samples demonstrate quite close distributions.
Our PRG candidates show approximately the same apparent magnitudes,
redshifts, and luminosities. Thus, we can consider our sample as a notable 
supplement to the sample of already known PRGs selected from the SDSS.

Fig.~\ref{char1}\textit{d} illustrates the standard observational selection effect -- 
among distant objects we choose only the brightest. The dashed lines in the figure 
distinguish two subsamples: nearby PRGs with $z \leq 0.03$ (the vertical line) 
and bright galaxies with $M_r \leq -21$ (the horisontal line). Both subsamples 
are almost not intersecting.

\begin{table}
\caption{Average characteristics of PRGs.}
 \label{tab2}
  \begin{tabular}{cccc}
  \hline 
  \hline
       & All      & Nearby         & Bright \\
       &          & $z \leq 0.03$ & $M_r \leq -21^m$ \\
\hline 
Number & 109             & 26               & 35 \\
$z$    & 0.055$\pm$0.034 & 0.019$\pm$0.009  & 0.073$\pm$0.039 \\
$M_r$  & --20.6$\pm$1.0  & --19.7$\pm$1.1   & --21.6$\pm$0.4  \\
$(g - r)_0$ & +0.73$\pm$0.15 & +0.72$\pm$0.20 & +0.74$\pm$0.12 \\
$i$ ($^\circ$) & 16.4$\pm$14.0 & 25.8$\pm$16.9 & 15.4$\pm$12.4 \\
$D_\mathrm{HG}$ (kpc) & 15.8$\pm$7.9   & 11.2$\pm$5.2  & 22.1$\pm$8.3 \\
$D_\mathrm{PS}$ (kpc) & 23.3$\pm$12.9  & 17.5$\pm$11.2 & 27.0$\pm$10.7 \\
$D_\mathrm{PS}$/D$_\mathrm{HG}$ & 1.60$\pm$0.89 & 1.74$\pm$1.18 & 1.32$\pm$0.64 \\
\hline\\
  \end{tabular}
   
\end{table} 

In Table~\ref{tab2} we compare average characteristics
of PRGs in three sets of data. Designation ``all'' in the table means
a joint sample (the present work + \citealt{smirmois2013}), the ``nearby'' and
``bright'' ones include only galaxies with $z \leq 0.03$ or $M_r \leq -21$, 
respectively, from the joint sample. As one can see, typical PRGs are rather bright
galaxies. Their average rest-frame integral color $g - r$ corresponds 
well to early-type (S0) galaxies (e.g. \citealt{fukug2007}). Polar
features are large-scale structures with typical sizes $\sim$20--30 kpc
and their orientation is close to orthogonal with respect to the host galaxies.
(The angle $i$ in Table~\ref{tab2} is an apparent angular distance away from 
perpendicularity of the host galaxy and the ring: $i = 90^\circ - \Delta$P.A.)
For the majority of PRGs, the diameter of the ring exceeds the
size of the host galaxy: >70\% of PRGs show the ratio $D_\mathrm{PS}/D_\mathrm{HG}$>1.

Effective radii of the sample galaxies reach several kiloparsecs 
(Table~\ref{tab1}), which is typical for bright early-type galaxies. In the
plane $M_r$--$R_{eff}$ the PRG candidates are located in the area where 
the characteristics of previously known PRGs are resided (see fig.\,4 in 
\citealt{rc2015}).

Our method of the surface brightness enhancement (Sect.~\ref{reduction}) allowed us to 
study faint structures around the PRG candidates. According to numerical
simulations, different formation mechanisms predict different external
morphology of PRGs. For instance, the merger scenario leads to a 
faint stellar halo surrounding the PRG, while the accretion scenario does 
not (\citealt{bourcomb2003}).
Our study of the PRG images shows that almost half of the objects (14 out of 31) 
can be surrounded by a weak ($\mu_r\sim26-26.5$~mag\,arcsec$^{-2}$) envelope (see 
several examples in Fig.~\ref{images}: 
SDSS\,010816.84-082317.6, SDSS\,041557.96-051655.8,  SDSS\,161103.94+140043.6,
and SDSS\,234407.32+253657.7).
This statistics is very preliminary and is based on insufficiently deep 
observational data. But it shows that the merger scenario can make
an appreciable contribution to the formation of PRGs. Future deep
photometric observations of PRGs should clarify this issue.

\subsection{The luminosity--size relation}

\begin{figure}
\includegraphics[width=11.0cm, angle=-90, clip=]{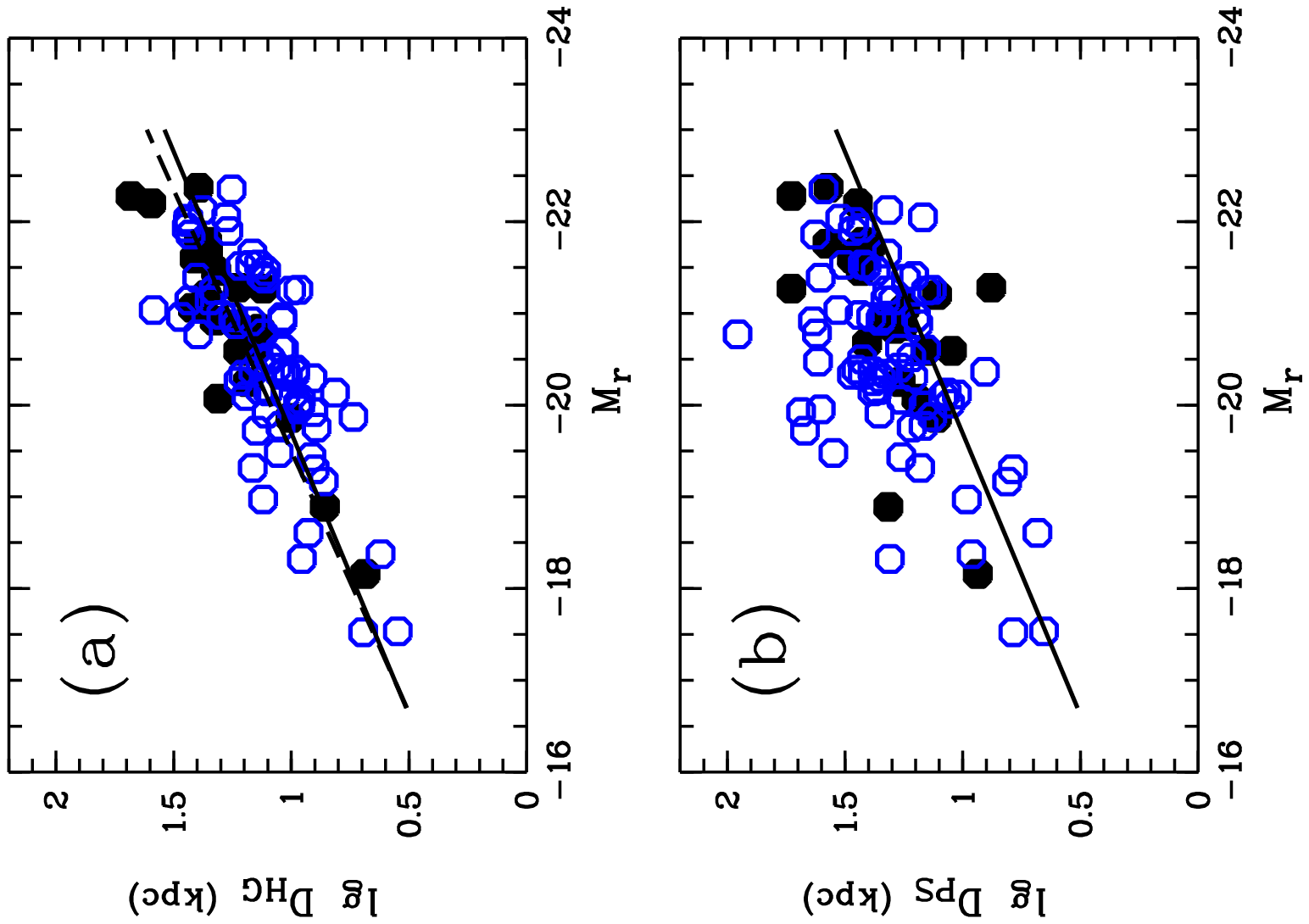}
\caption{Absolute magnitude of the PRG vs. (a) diameter of the HG or
(b) diameter of the PS (the solid circles -- the data from Table~\ref{tab1},
the open circles -- the characteristics of PRGs from \citealt{smirmois2013}).
The dashed line presents the relation for nearby galaxies according to
\citet{bergh2008}. The solid lines in both figures show a linear 
approximation for the central galaxies of PRGs.}
\label{sizes}
\end{figure}

In Fig.~\ref{sizes}, we show the luminosity -- size relation for the host galaxies
and for the polar structures. As one can see, the central objects of PRGs show a 
tight correlation between the luminosity and diameter (Fig.~\ref{sizes}\textit{a}). 
The best fitting
line for our joint sample is $\log$\,$D_\mathrm{HG}$ = --0.163\,$M_r$ -- 2.21 (the solid
line in the figure). The PRGs follow this relation with a r.m.s. scatter 
of 0.13. The dashed line in Fig.~\ref{sizes}\textit{a} is the average relation for 
nearby galaxies of all morphological types (\citealt{bergh2008}) shifted from the $B$  
passband to the $r$ according to \citet{cook2014}.
Interestingly, HGs of PRGs follow this relation well. 

The same relation but for the diameter of the polar structure is shown in 
Fig.~\ref{sizes}\textit{b}. The PSs follow the relation with approximately 
the same angle but it is shifted to higher sizes. At a given total luminosity of 
PRGs, polar rings are, on average, larger in comparison with their host galaxies. 
It is a natural result, since the direct comparison of the two subsystems gives
$D_\mathrm{PS}/D_\mathrm{HG} > 1$ (see Table~\ref{tab2}). 

More interestingly, the PSs
display a notable larger scatter around the average trend (r.m.s. dispersion is
0.20) compared to the hosts (Fig.~\ref{sizes}\textit{a}). The increasing scatter can be
a direct consequence of a secondary origin of polar structures. Polar
features are relatively young and they are formed during non-stationary external 
processes (accretion, merging). Therefore, at an early stage of formation, 
these structures can be asymmetric and perturbed (see, e.g., \citealt{brook2008}
simulations.) In the course of the subsequent evolution 
the secondary subsystems are likely to settle into a narrower luminosity -- size
relation.

\subsection{The distribution of angles between the rings and the central galaxies}

\begin{figure}
\includegraphics[width=8.0cm, angle=-90, clip=]{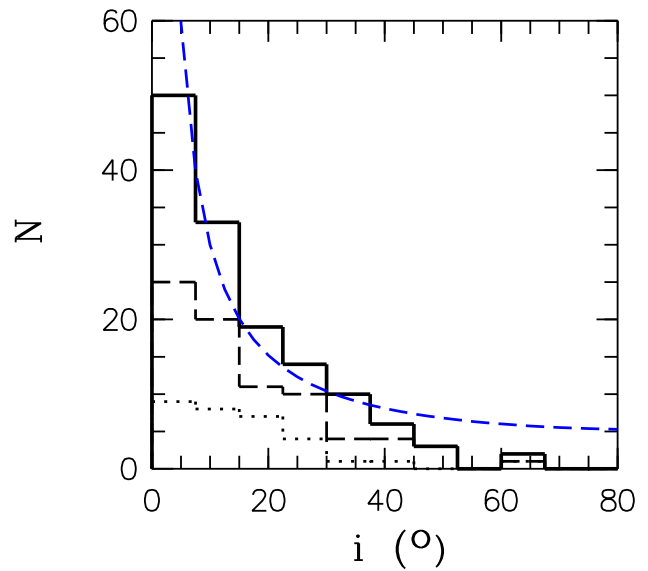}
\caption{The distribution of PRGs as a function of the angle between 
the PS and the perpendicular to the disc of the central galaxy. The dotted
line shows the data for our sample (Table~\ref{tab1}), the dashed histogram -- 
the \citet{smirmois2013} measurements, the solid line -- all available data.
The dashed curved line illustrates the arbitrary scaled dependence 
$\propto$1/sin\,$i$ (see the text for details). }
\label{incl}
\end{figure}

Previous attempts to investigate statistics of apparent angles between the rings 
and the host galaxies were based on relatively small samples (\citealt{whit1991},
\citealt{rc2015}). (We prefer to investigate the apparent angle between 
these two subsystems, rather than to analyse the spatial angle, as done in 
\citealt{smirmois2013}, since in their case we have to know the 
intrinsic flattening and the shape of the two components.) But now we
can consider much more data. In Fig.~\ref{incl} we summarise the data from 
Table~\ref{tab1} (31 galaxies, the dotted line) and from \citet{smirmois2013} 
(78 PRGs, the dashed line). Also, we added 28 measurements from
\citet{whit1991} and constructed the final distribution for 137 PRGs
(the solid histogram). (Our final sample includes, thus, 137 original objects, 
without duplicates.)

As one can see in Fig.~\ref{incl}, the majority of PRGs, indeed, show almost orthogonal
outer structures. From the combined sample of 137 objects we conclude that
50\% of the PRGs demonstrate $i \leq 10^{\circ}$, 75\% -- $i \leq 23^{\circ}$,
and 90\% -- $i \leq 35^{\circ}$. If we consider a relatively complete subsample
with $z \leq 0.03$ (26 PRGs from the joint sample 
Table~\ref{tab1}  + \citealt{smirmois2013} and 12 
galaxies from \citealt{whit1991}), the distribution becomes wider: 
50\% of the PRGs with $i \leq 17^{\circ}$, 75\% -- $i \leq 33^{\circ}$, and 90\% --
$i \leq 41^{\circ}$.

Interpretation of the distribution by $i$ is not a trivial task since
it is not a true angle between the central galaxy plane and the plane of the PS 
but merely an apparent, projected angle. Moreover, the distribution in
Fig.~\ref{incl} is strongly affected by the observational selection. 

For illustrative purposes, we placed in Fig.~\ref{incl} the expected dependence
of the settling time for a gaseous ring on its inclination to the
preferred, stable plane (the blue dashed curve). This settling time
is inversely proportional to cosine of an angle by which the ring is tilted out of
a stable plane (e.g. \citealt{tohline1990}). If we assume that the plane 
of the HG defines the preferred plane, this dependence
transforms to 1/$\sin$\,$i$. It is evident from the figure that the observed
distribution does approximately follow this law.  For the rings with
small $i$ (nearly polar), the settling time can be long and we can observe 
a lot of such systems in the Universe. For more inclined structures, the rings
ought to settle relatively rapidly and we have less chances to observe such objects.

\section{CONCLUSIONS}

In our work we presented a new sample of 31 candidates for polar-ring galaxies. 
The galaxies were selected on the basis of their morphology from the SDSS.
Our list of PRGs is a valuable addition to the SPRC catalogue (Fig.~\ref{char1}). 

From the analysis of general characteristics of PRGs we derived the following 
main conclusions:

-- The host galaxies of PRGs follow the luminosity--size relation for normal
galaxies. The polar structures show similar relation but with a larger 
scatter (Fig.~\ref{sizes}). This enhanced scatter may be a consequence of a secondary 
origin of polar rings.

-- A nearly polar orientation dominates among the rings of PRGs: approximately
half of all PRGs demonstrate extended structures within 20$^{\circ}$ 
from perpendicularity (Fig.~\ref{incl}).

On the basis of the surface brightness enhancement we tried to detect
faint structures around the PRG candidates. The obtained results are
uncertain (Sect.~\ref{char}). Possible existence of outer stellar envelopes
around PRGs is a subject of our future study.

A gradual increase of discovered polar-ring galaxies makes it possible to 
pose new interesting questions about the nature and origin of these unique objects. 
One can hope that these galaxies will be put soon into the general 
picture of the formation and evolution of galaxies.

\section*{Acknowledgements}

We thank anonymous referee for helpful constructive comments that improved 
the paper.

Aleksandr Mosenkov expresses gratitude for the grant of the Russian Foundation for 
Basic Researches number mol\_a 18-32-00194.

This research has made use of the NASA/IPAC Extragalactic Database (NED) 
which is operated by the Jet Propulsion Laboratory, California Institute 
of Technology, under contract with the National Aeronautics and Space 
Administration.

Funding for the SDSS has been provided by the Alfred
P. Sloan Foundation, the Participating Institutions, the
National Science Foundation, the U.S. Department of Energy,
the National Aeronautics and Space Administration, the Japanese 
Monbukagakusho, the Max Planck Society, and
the Higher Education Funding Council for England. The
SDSS Web Site is http://www.sdss.org/.

\appendix

\section{Reduced galaxy images}
\label{red_images}

\begin{figure*}
\centering
\includegraphics[width=18cm, angle=0, clip=]{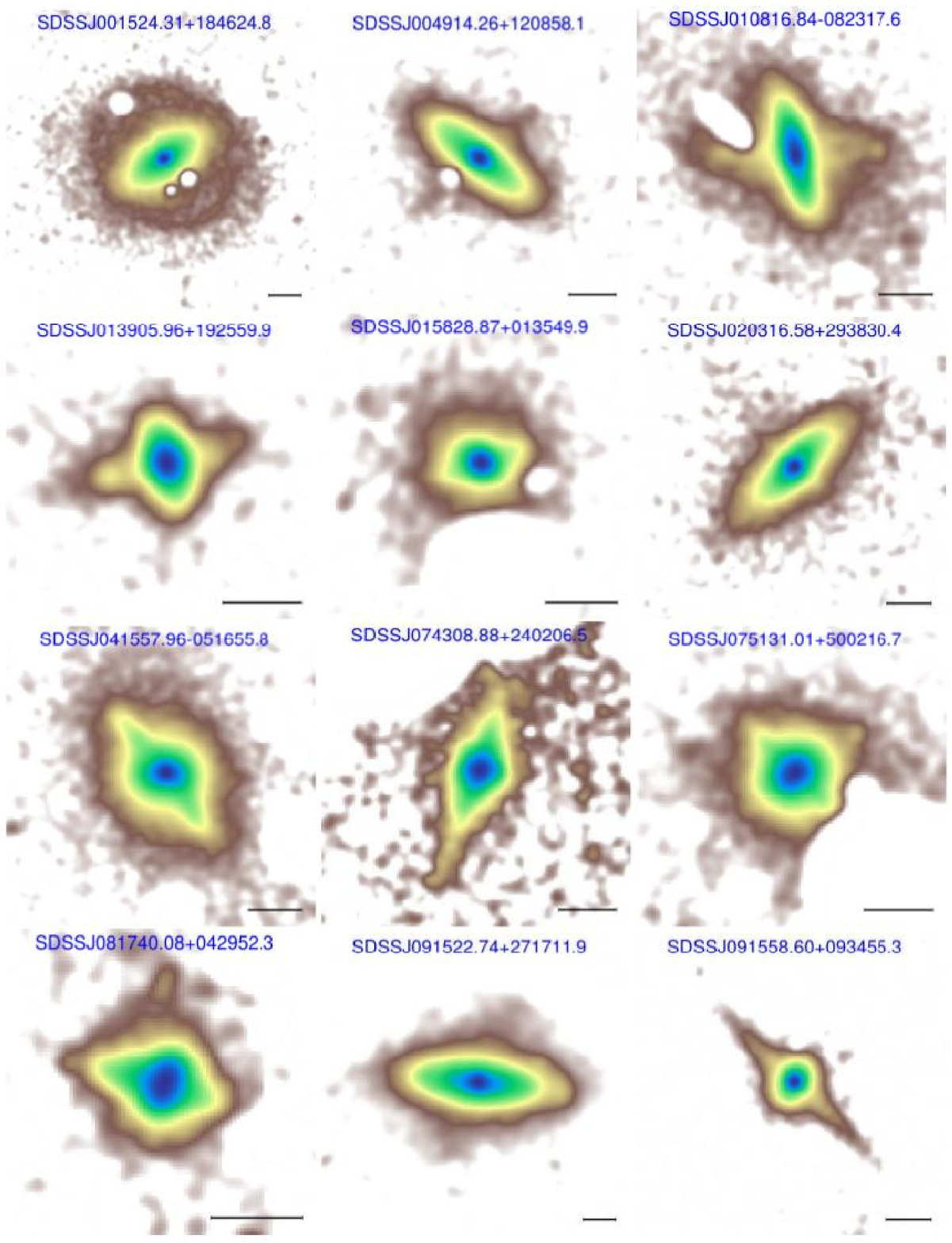}
\caption{Reduced (stacked) $gri$-images of the candidates to PRGs. 
The contaminating sources 
have been masked. The scale bar in each plot corresponds to 10\arcsec.}
\label{images}
\end{figure*}

\addtocounter{figure}{-1}
\begin{figure*}
\centering
\includegraphics[width=18cm, angle=0, clip=]{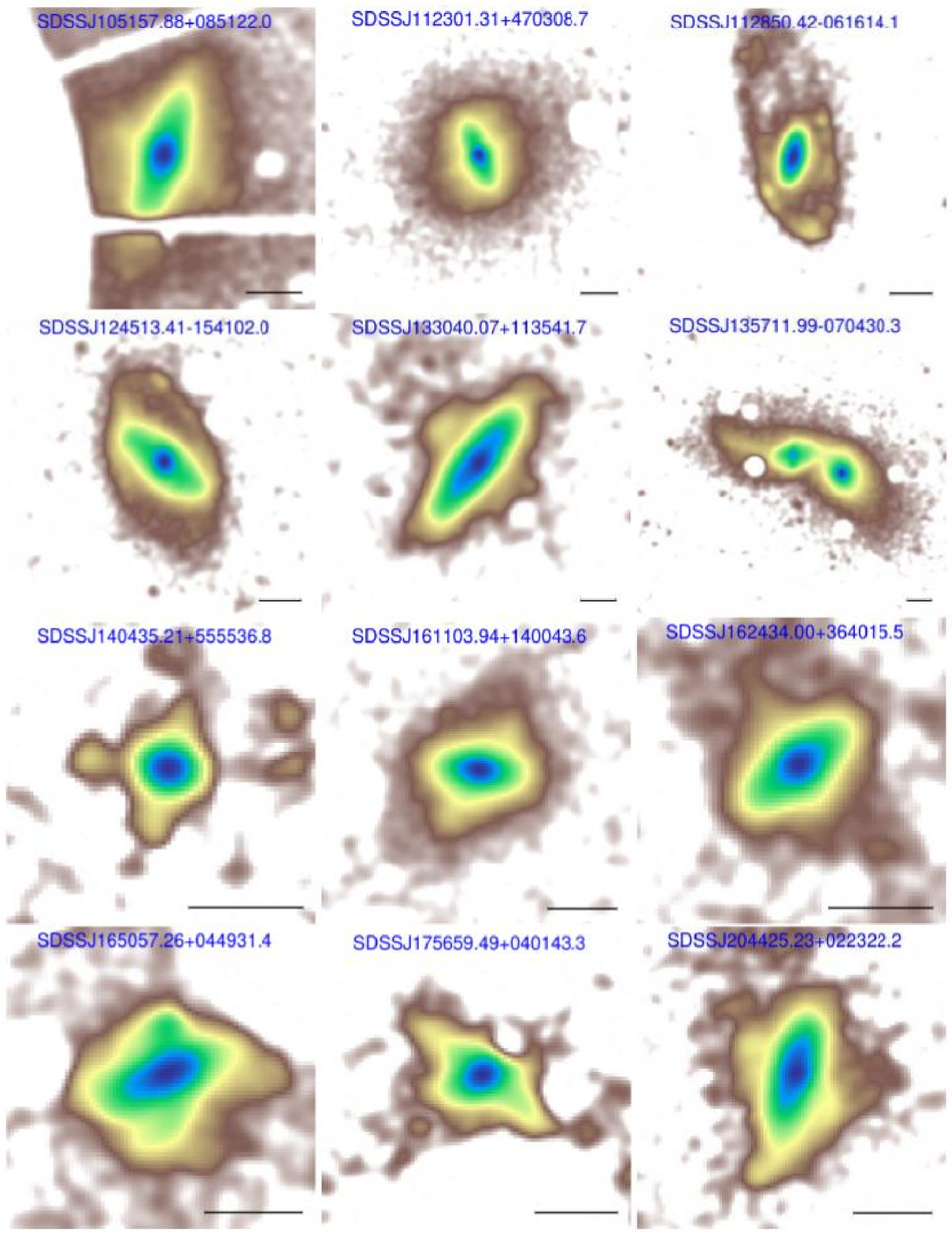}
\caption{(continued)}
\end{figure*}

\addtocounter{figure}{-1}
\begin{figure*}
\centering
\includegraphics[width=18cm, angle=0, clip=]{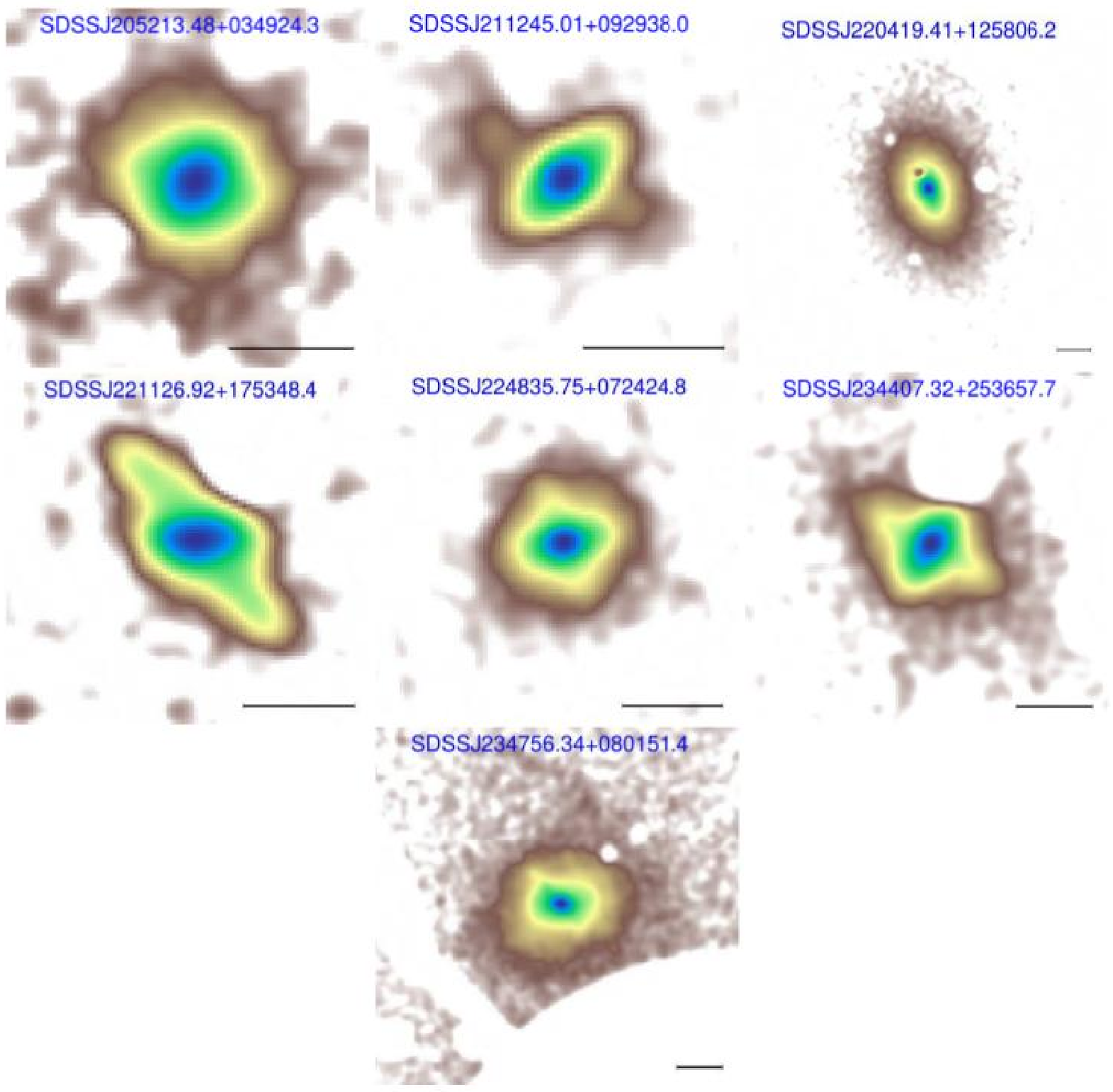}
\caption{(continued)}
\end{figure*}



\bibliographystyle{mnras}
\bibliography{art}

\label{lastpage}
\end{document}